\documentclass[aps,english,prl,twocolumn,superscriptaddress]{revtex4}
\usepackage{graphicx}

\begin{document}

\title{New Phases of Germanene}

\author{V. Ongun \"{O}z\c{c}elik}
\affiliation{UNAM-National Nanotechnology Research Center, Bilkent University, 06800 Ankara, Turkey}
\affiliation{Institute of Materials Science and Nanotechnology, Bilkent University, Ankara 06800, Turkey}
\author{E. Durgun}
\affiliation{UNAM-National Nanotechnology Research Center, Bilkent University, 06800 Ankara, Turkey}
\affiliation{Institute of Materials Science and Nanotechnology, Bilkent University, Ankara 06800, Turkey}
\author{S. Ciraci}
\affiliation{Department of Physics, Bilkent University, Ankara 06800, Turkey}

\begin{abstract}
Germanene, a graphene like single layer structure of Ge, has been shown to be stable and recently grown on Pt and Au substrates. We show that a Ge adatom adsorbed to germanene pushes down the host Ge atom underneath and forms a dumbbell structure. This exothermic process occurs spontaneously. The attractive dumbbell-dumbbell interaction favors high coverage of dumbbells. This letter heralds stable new phases of germanene, which are constructed from periodically repeating coverage of dumbbell structures and display diversity of electronic and magnetic properties.
\end{abstract}

\maketitle

Three dimensional (3D) layered bulk phases, such as graphite,\cite{geim} BN\cite{novo} and MoS$_2$\cite{mos2} have led to the synthesis of single layer, honeycomb structures of those materials, which were initially conjectured to be unstable.\cite{coleman,naturenano,nature499,nicolosi} The existence of single layer, graphene like structures of other Group IV elements, like Si and Ge, have been ruled out since these elements do not have layered allotropes like graphite that would allow the synthesis of their single layer structures. Surprisingly, based on state of the art first-principles calculations, silicene,\cite{takeda,engin} germanene,\cite{seymur1,ansiklo} most of III-V and II-VI compounds\cite{engin,ansiklo} and several transition metal dichalcogenides and oxides\cite{can} have been shown to form stable, single layer honeycomb structures. Moreover, it has been also shown that silicene and germanene share several of the exceptional properties of graphene, such as $\pi$- and $\pi^*$-bands linearly crossing at the Fermi level and hence forming Dirac cones, the ambipolar effect and the family behavior observed in nanoribbons.\cite{seymur1,seymur2} Advances in growth techniques have enabled the synthesis of some of these predicted single layer structures, in particular the growth of single and multilayer silicene on Ag(111) substrates\cite{lelay,lelay2,lelay3} and the growth of germanene on Pt and Au substrates\cite{germanene,lelay4,germanane} were recently succeeded.

\begin{figure}
\includegraphics[width=8cm]{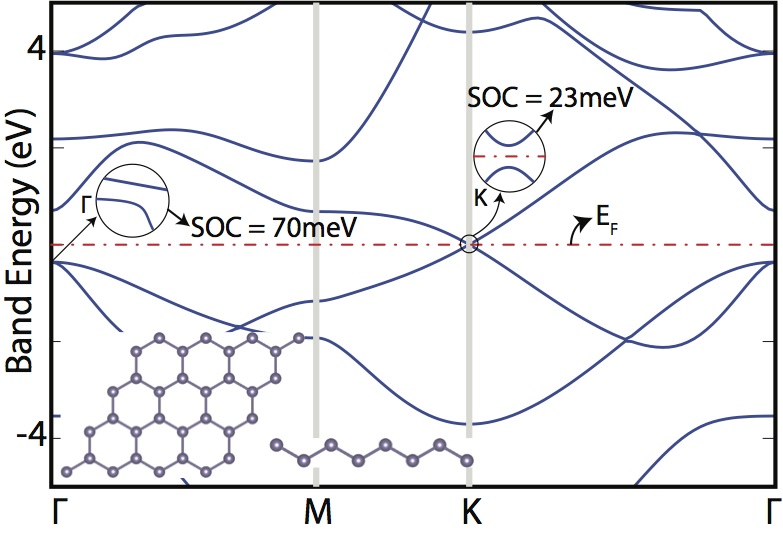}
\caption{Electronic structure of monolayer germanene. The splittings due to the spin-orbit coupling (SOC) are shown in the magnified insets. The Fermi level is set to the zero of energy and is indicated by the dashed-dotted line. The atomic structure of buckled germanene is illustrated by inset.}
\label{fig1}
\end{figure}

\begin{figure}
\includegraphics[width=8cm]{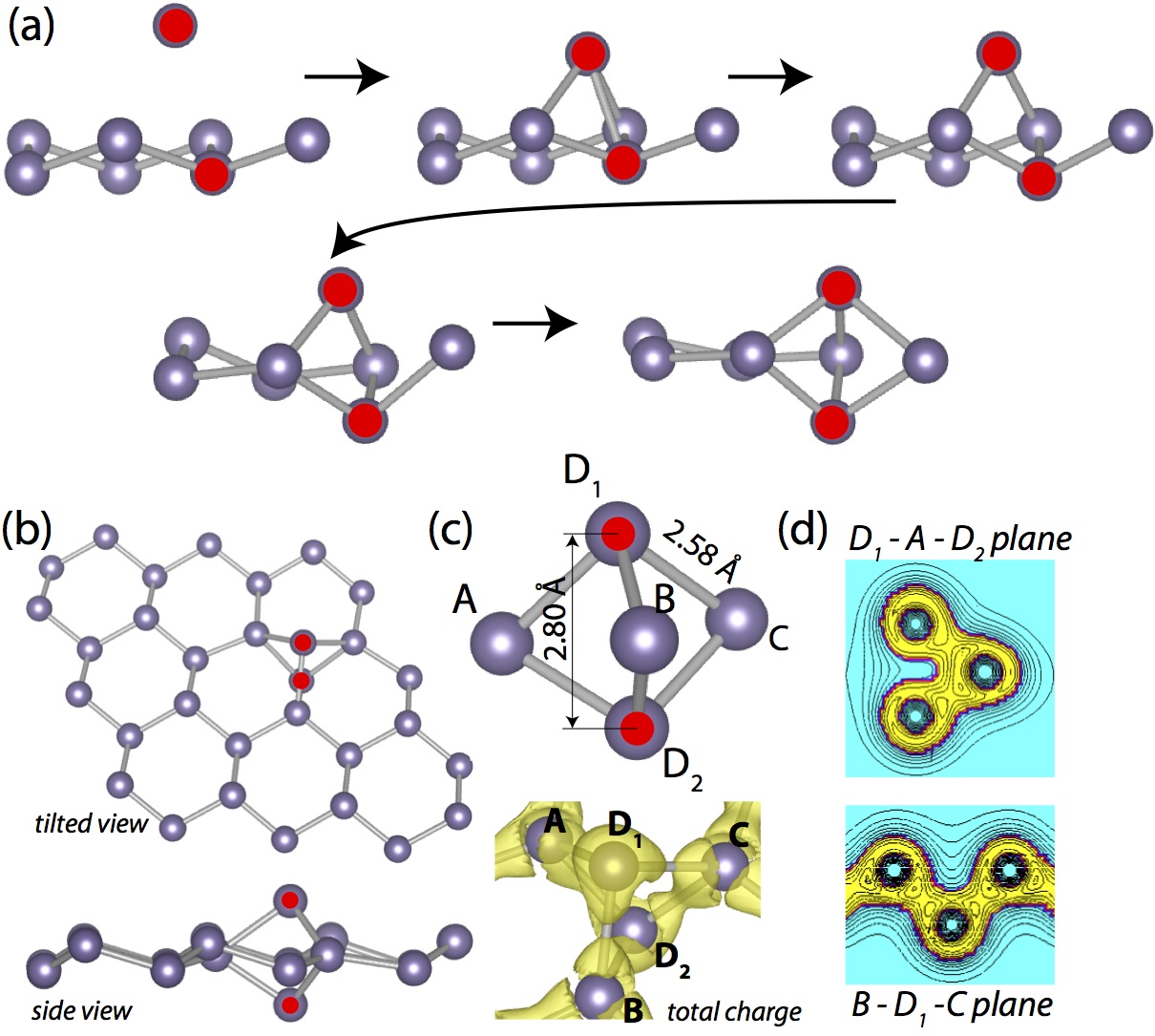}
\caption{(a) Snapshots of conjugate gradient steps in the course of the formation of a dumbbell (DB) structure. The external Ge adatom first approaches to the germanene layer from the top site, and eventually constructs the DB structure by pushing the host Ge atom down.  (b) Top and side views of DB formed on ($4 \times 4$) germanene. Two Ge atoms of DB are highlighted by red. (c) The DB zoomed in along with the total charge density isosurfaces. (d) Contour plots of the total charge density on planes passing through $D_1-A-D_2$ and $B-D_2-C$ atoms. Note that although the DB atoms make bonds with nearest germanene atoms, there is no bonding between the DB atoms, $D_1$ and $D_2$. }
\label{fig2}
\end{figure}

In this letter, we find that a single Ge adatom adsorbed to germanene constructs a dumbbell (DB) structure. Even more remarkable is that new phases can be constructed by the periodic coverage of germanene with DBs. Depending on the coverage of DBs, these stable phases can be metal or narrow band gap semiconductor in magnetic or nonmagnetic states and hence they attribute diverse functionalities to germanene. We believe that the intriguing stacking of these phases can pave the way to the layered phase of bulk germanium, i.e. germanite.

Our predictions are obtained from the state of the art first-principles pseudopotential calculations based on the spin-polarized Density Functional Theory (DFT) within generalized gradient approximation(GGA) including van der Waals corrections.\cite{grimme06} We used projector-augmented wave potentials (PAW)\cite{blochl94} and the exchange-correlation potential is approximated with Perdew-Burke-Ernzerhof (PBE) functional.\cite{pbe} The Brillouin zone (BZ) was sampled in the Monkhorst-Pack scheme, where the convergence in energy as a function of the number of \textbf{k}-points was tested. The \textbf{k}-point sampling of (21$\times$21$\times$1) was found to be suitable for the BZ corresponding to the primitive unit cell of germanene. For larger supercells this sampling has been scaled accordingly. The number of \textbf{k}-points were further increased to (48$\times$48$\times$1) in the density of states calculation. Atomic positions were optimized using the conjugate gradient method, where the total energy and atomic forces were minimized. The energy convergence value between two consecutive steps was chosen as $10^{-5}$ eV. A maximum force of 0.002 eV/\AA~ was allowed on each atom. Numerical calculations were carried out using the VASP software.\cite{vasp} Since the band gaps are underestimated by DFT, we also carried out calculations using the HSE06 hybrid functional\cite{hse}, which is constructed by mixing 25\% of the Fock exchange with 75\% of the PBE exchange and 100\% of the PBE correlation energy. The phonon dispersion curves were calculated using small displacement method.\cite{alfe2009phon}

It is known that the splitting of threefold valence band at the center of the Brillouin zone due to spin-orbit coupling is significant (290 meV) for Ge crystal in cubic diamond (cdGe) structure.\cite{kittel} The extend of spin-orbit splitting in germanene is crucial as a starting point of the present study. Despite 1 meV spin-orbit splitting in silicene,\cite{socsi} the spin-orbit splitting of germanene at $\Gamma$- and K-points of the hexagonal Brillouin zone are calculated to be 70 meV and 23 meV. Accordingly, it is set that at the K, K$^{'}$-points, where the linear bands cross at the Fermi level in the absence of spin-orbit coupling, a gap of 23 meV is opened. This way the semimetallic state and massless Fermion behavior of electrons of pristine germanene are discarded as seen in ~\ref{fig1}. This result is critical for our study dealing with new derivatives of pristine germanene.

Having clarified the effect of spin-orbit coupling on pristine germanene, we next examine the construction of a single DB structure. In ~\ref{fig2}(a), we present various stages of conjugate gradient calculations taking place in the course of the adsorption of single Ge adatom. In the presence of an external and free Ge adatom, the formation of DB structure on germanene is spontaneous. The external Ge adatom eventually moves closer to the germanene surface and makes a bridge bond with two underlying Ge atoms of germanene. It then starts to push the Ge atom underneath further down until the final DB forms as shown in ~\ref{fig2} (a). As a concerted process, two Ge atoms above and below the germanene surface, named as D$_1$ and D$_2$, donate significant electronic charge to the three nearest Ge atoms of germanene and hence each forms three strong Ge-Ge bonds with a length of 2.58 \AA. With these additional bonds with DB, these three Ge atoms of germanene become fourfold coordinated. D$_1$ and D$_2$ by themselves engage in a 3+1 coordination, since each has three nearest neighbor Ge atoms at a distance of 2.58 \AA. Whereas the D$_1$ - D$_2$ distance of 2.80 \AA~is slightly larger. At the end, the resulting DB corresponds to a local minimum on the Born-Oppenheimer surface and remains stable. As for the DB decorated germanene, it is a structure between the fourfold coordinated cdGe and the three fold coordinated single layer, buckled germanene. Since moving a DB from one place to another by breaking three Ge-Ge bonds involves an energy barrier, any pattern of DBs on germanene is expected to remain stable. It is noted that DB has been found to be the second most energetic defect structure of carbon adatom on graphene.\cite{can} Later, it has been found to be the most energetic defect structure of Si adatom on silicene,\cite{ongun1, kaltsas} and has been demonstrated that it can form spontaneously without an energy barrier as long as a Si adatom on silicene is available.\cite{ongun2} In the present paper, we demonstrated that Ge DB can form new phases (or derivatives) of germanene with novel electronic and magnetic properties.

Since the construction of a single DB is an exothermic process and hence does not involve any energy barrier, the formation of DB structure is unavoidable as long as a free Ge adatom is present at the close proximity of the surface. We define the associated binding energy as $E_b=E_{T}[germanene]+ E_T[Ge] - E_{T}[germanene+DB]$; in terms of the total energies of germanene+DB, pristine germanene and free Ge adatom. The binding energies are calculated for a single DB in the ($n \times n$) hexagonal supercells with varying values of $n$. Accordingly, $E_{b}$ is the energy gained from the construction of a single DB through the adsorption of a single Ge adatom to germanene and $E_{b} >$0 indicates an exothermic process. For an isolated DB calculated in a large supercell with $n$=8, $E_{b} \sim$ 3.4 eV; but it increases with decreasing $n$ or decreasing DB-DB distance due to the attractive interaction among DBs as discussed in the next paragraph. In ~\ref{fig2} (b) and (c), the atomic configurations of a single DB and its relevant structural parameters together with isosurfaces of charge density of Ge-Ge bonds around DB are shown. The charge density counter plots calculated on various planes are presented in ~\ref{fig2}(d). The bonding of the DB atoms (D$_1$ and D$_2$) with the nearest Ge atoms of germanene are clearly seen. Notably, there is no bonding between D$_1$ and D$_2$.

\begin{figure}
\begin{center}
\includegraphics[width=8cm]{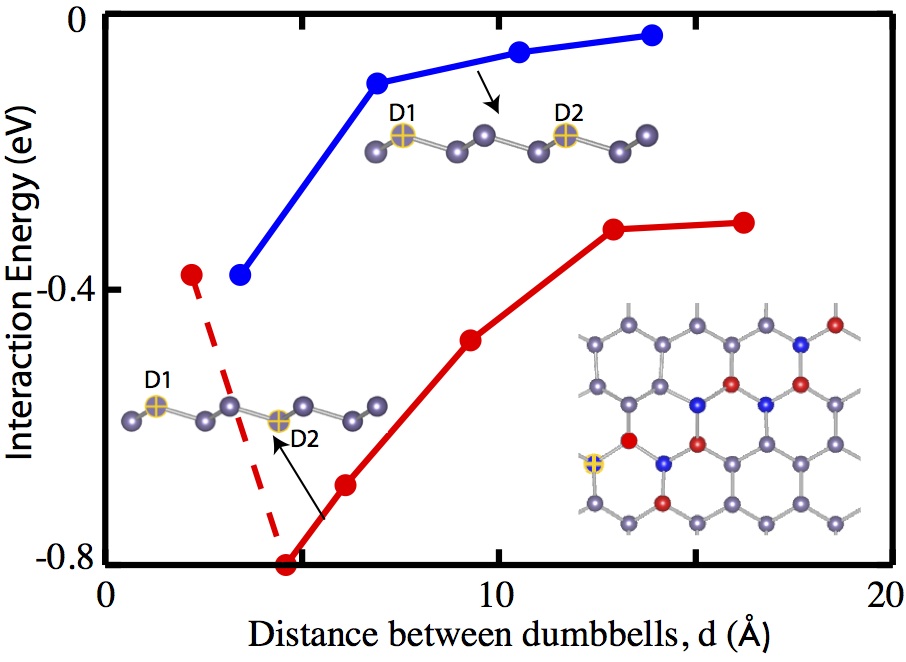}
\caption{The interaction energy versus the distance between two DBs, $d$ on the ($8 \times 8$) supercell of germanene. The blue and red curves represent the variation of interaction energies for DBs formed on sites with the same and opposite bucklings, respectively. The interaction energy between two DBs situated at the same buckling is set to zero for large $d$. Negative energy indicates attractive interaction. One DB is permanently present on the yellow site and the second DB is placed on various positions shown by the blue and red marks in the inset. The attractive interaction energy falls suddenly when the second DB following the red path is situated at the nearest neighbor distance to the first DB. }
\label{fig3}
\end{center}
\end{figure}

The DB-DB interaction on the surface of germanene is crucial for the growth of germanene phases. While an attractive interaction between two DBs can lead to the domain structure, a repulsive interaction at small DB-DB distance $d$ is expected to favor phases with uniform coverage of DB. Therefore, we next investigate what happens if a Ge adatom is introduced in addition to an existing DB. Our calculations show that rather than bonding to D$_1$ or D$_2$ dumbbell atoms and forming a short Ge-Ge chain on top of them, Ge adatom migrates on germanene substrate and forms another DB. Hence, each Ge adatom introduced to the surface of germanene favors the construction of a new DB, as long as a proper position is available. The DB-DB coupling is calculated by placing two DBs in an ($8 \times 8$) supercell; one at a fixed corner, and the second one placed at different sites on the supercell as described in ~\ref{fig3}. For each lateral position of the second DB, the coordinates of rest of the atoms including the height of the second DB are fully optimized. Apparently, an attractive interaction is set even for the large DB-DB distance. The DBs tend to be close to each other and hence to form domains. However, as long as germanene continues to be fed by Ge adatoms, domains join to form full coverage. Here one distinguishes two classes of sites for the second DB: One class of sites is the high buckled sites of germanene, the other class is the low buckled sites. It should be noted that low and high buckled sites are equivalent if there is only one DB due to the upside and downside symmetry. The site-specific variation of the total energy is shown in ~\ref{fig3}, where the interaction energy versus $d$ curve follow different paths. Accordingly, the formation of two DBs on sites with opposite bucklings appears to be more favorable energetically, by $\sim$0.2-0.3 eV, since the gain of energy through the buckling of bare germanene is preserved. However, this difference of energy diminish as $d \rightarrow \infty$.  Notably, it is not possible to create two DBs on the nearest neighbor sites of the lattice. This situation results in the constrained structure optimization in ~\ref{fig3} (shown by the dashed line) as a sudden fall of the attractive interaction energy when the Ge adatom is placed at the first nearest neighbor of the existing DB. The second DB tends to move to the second nearest neighbor position when the constraints are lifted. If the second DB is situated at the second nearest neighbor distance from the first DB (where both DBs are situated on the sites of same buckling), Ge atoms of germanene at the first nearest neighbor distance to these two DBs become five fold coordinated.

\begin{table*}
\caption{Calculated values for the various phases of germanene+DB, where DBs form periodically repeating supercells on germanene with 2D hexagonal or rectangular lattice. 2D Lattice: H hexagonal or R rectangular; Mesh: ($m \times n$) cell in terms of the primitive hexagonal or rectangular unit cell of germanene; $N$: Number of Ge atoms (including DB) in each supercell; $d$: shortest distance between two DBs; $A$: the area of the supercell; $\mu$: magnetic moment per supercell;  $ES$: Electronic structure specified as metal M, or semiconductor with the band gap between valance and conduction bands calculated by GGA and HSE (for the spin polarized cases the gap between spin up - spin up and spin up - spin down bands are shown); $E_b$: Binding energy per Ge adatom relative to germanene or average binding energy if there is two DB in each cell; $E_{C}$: Cohesive energy (per atom) of Ge atom in Germanene+DB phase; $E^{s}_{C}$: Cohesive energy per area; $\Delta E_{C}$: difference between the cohesive energies of a Ge atom in Germanene+DB and in pristine germanene, where positive values indicates that germanene+DB phase is favorable. For bare germanene $E_{C}=$3.39 eV/atom. TDP, HDP, RDP and DHP are described in the text as well as in Supplement.}
\label{table1}
\begin{center}
\resizebox{\columnwidth}{25mm}{
\begin{tabular}{cccccccccccc}
\hline  \hline
Lattice & Mesh & $N$ & $d$(\AA) & $A$(\AA$^2$) & $\mu$($\mu_B$)  & $ES_{GGA}$ (eV)& $ES_{HSE}$ (eV) &  $E_b$ (eV)  & $E_{C}$ (eV) & $E^{s}_{C}$ & $\Delta E_{C}$ (eV)  \\
\hline
H/FDP & $1 \times 1$ & 3 & 3.77 & 12.34 & 0 & M & M & 4.00 & 3.60 & 0.873 & 0.204 \\
R/RDP & $1 \times 1$ & 5 & 3.39 & 24.21 & 0 & M & M & 3.73 & 3.46 & 0.715 & 0.069 \\
H/TDP & $\sqrt{3} \times \sqrt{3}$ & 7 & 6.84 & 40.57 & 0 & M & M & 3.75 & 3.44 & 0.595 & 0.052 \\
H/HDP & $\sqrt{3} \times \sqrt{3}$ & 8 & 3.85 & 38.60 & 0 & 0.53 & 0.73 & 4.03 & 3.55 & 0.736 & 0.160 \\
H & $2 \times 2$ & 9 & 7.79 & 52.54 & 0 & 0.16 & 0.32 & 3.72 & 3.43 & 0.585 & 0.037 \\
H/DHP & $2 \times 2$ & 10 & 4.50 & 52.54 & 0 & M & 0.46 & 4.39 & 3.59 & 0.680 & 0.198 \\
H & $3 \times 3$ & 19 & 11.68 & 118.22 & 0 & 0.29 & 0.63 & 3.54 & 3.40 & 0.551 & 0.008 \\
H & $4 \times 4$ & 33 & 16.06 & 223.36 & 2 & 0.42 / 0.06 & 0.77/0.44 & 3.44 & 3.39 & 0.495 & 0.002 \\
R & $2 \times 1$ &  9 &  6.95 &  48.43 & 0 & M & M & 3.85 & 3.48 & 0.189 & 0.189 \\
H & $5 \times 5$ & 51 & 19.87 & 342.05 & 2 & 0.33 / 0.03 & 0.61/0.36 & 3.40 & 3.39 & 0.510 & 0.000 \\

\hline
\hline
\end{tabular}
}
\end{center}
\end{table*}

In the case of uniform and periodic coverage, DBs form a mesh or periodically repeating supercells on germanene. The properties of the resulting germanene+DB phases depend on the size and geometry of supercells constructed from ($n \times n$) hexagonal primitive unit cells or ($m \times n$) rectangular unit cells of germanene, as well as the number of DBs in each supercell or the DB-DB distance, $d$. In ~\ref{table1}, the energetics and the relevant data of the selected phases having 2D hexagonal or rectangular lattice structures are presented. The cohesive energy ($E_{C}=E_{T}[Ge] - E_{T}[germanene+DB]/N$,\cite{kittel} which is the difference between the energy of one free Ge atom and the energy of germanene+DB phase per atom or simply the energy gained per atom by constructing a particular germanene+DB phase), and cohesive energy per unit area (i.e. $E^{s}_{C}= N E_{C}/A$) are relevant measures for energetics of the phases. In particular, $E^{s}_{C}$ is a prime criterion which decides the phase that will grow on bare germanene. Here $N$ is the total number of Ge atoms in the supercell of a given phase and $A$ is the area of the supercell. Since each DB constructed on germanene lowers the energy, FDP (Full Dumbbell Phase) which corresponds to full coverage, attains the highest $E_C$, $E^{s}_{C}$ and $\Delta E_{C}$ among other phases listed in ~\ref{table1}. However, one Ge atom in each cell of FDP is forced to be six fold coordinated, and hence FDP is prone to structural instability. In fact, our ab-initio phonon calculations of this phase have branches with imaginary frequencies, which indicate structural instability as shown in the supplement. Another structure, RDP (Rectangular Dumbbell Phase) in ~\ref{table1} is also found to be unstable based on ab-initio phonon calculations. DHP (Double Hexagonal Phase), where the hexagons of germanene are nested by large DB hexagons has the highest $E_b$ and $E_C$ among the other stable phases listed in ~\ref{table1}. Additionally, two other phases, TDP (Trigonal Dumbbell Phase) and HDP (Hexagonal Dumbbell Phase) which have ($\sqrt{3} \times \sqrt{3}$) unitcells are of particular interest, since silicene grown on Ag(111) substrate also showed a ($\sqrt{3} \times \sqrt{3}$) pattern.\cite{lelay3,seymur3} HDP has two DBs per cell, such that DBs are situated at the corners of hexagons to form a honeycomb pattern. It appears that HDP having the maximum cohesive energy per area, $E^{s}_{C}=$ 0.735 eV/\AA$^2$ among the other stable phases listed in ~\ref{table1} is energetically the most favorable structure to grow on bare germanene.

The germanene+DB phases acquire permanent magnetic moments when $d > \sim$ 15 \AA, where the DB-DB coupling recedes and the DB behaves as a local defect on the germanene substrate with a total magnetic moment of $2 \mu_B$ per cell. We performed additional tests to understand the magnetic order of the system. To this extend, four DB structures were created on an ($8 \times 8$) supercell and the ferromagnetic, antiferromagnetic and paramagnetic states were investigated. For the ferromagnetic case, all DB atoms were given an initial spin in the same direction; for the anti-ferromagnetic case opposite spins were assigned to the adjacent dumbbells. Our results showed that the ferromagnetic ordering has the lowest total energy indicating that it is the most favorable configuration.

The energy difference between the ferromagnetic state and the antiferromagnetic state is 0.63 eV per supercell. Notably, upon the relaxation of magnetic states, the final magnetic state of the structure is found to be always ferromagnetic no matter what the initial direction of spins were. The energetics of various germanene+DB phases display interesting trends: In general, the cohesive energy increases with increasing DB coverage, which confirms the situation in ~\ref{fig3}. The binding energies, $E_b$, also show the same trend except for DHP. The energy values presented in ~\ref{table1} imply that the higher the DB coverage of a phase is, the higher its total energy gets.

\begin{figure*}
\includegraphics[width=14cm]{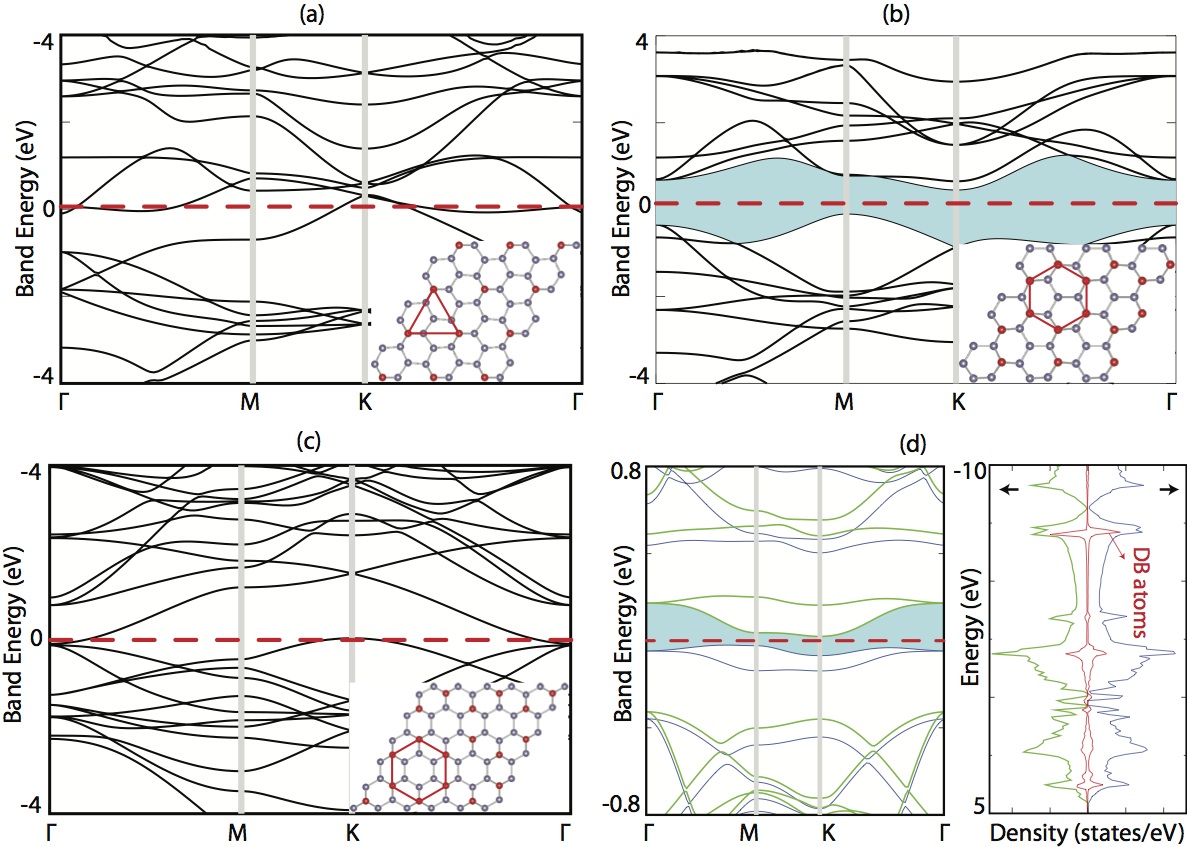}
\caption{ Electronic band structures of different phases of germanene. (a) TDP. (b) HDP. (c) DHP. (d) The triangular structure with DBs forming hexagonal ($4 \times 4$) supercells, where the total density of states are also shown. The spin up and spin down bands are shown in blue and green lines, respectively. The density of states projected to the DB atoms shown in red and are augmented three times for a better view. Atomic structures of TDP, HDP and DHP are described in the text as well as in Supplement.}
\label{fig4}
\end{figure*}

Dumbbells also modify the physical properties of germanene. In particular, the electronic and magnetic properties of  pristine germanene show dramatic changes depending on the coverage (see also ~\ref{table1}). Firstly, the spin-orbit coupling gives rise to significant splitting in the bands of germanene-DB phases. For example, the degenerate bands at the top of the valance band of ($4 \times 4$) mesh split by 12 meV. Normally, germamene+DB phases maintain their metallic state at high coverage, but they transform into semiconductor as the size of their cell($n$) or the DB-DB distance($d$) increases; with the exception of HDP. In ~\ref{fig4} we present the energy band structures of four different germanene+DB phases having hexagonal lattice. These are (i) TDP; (ii) HDP; (iii) DHP and (iv) triangular structure, where DBs form ($4\times4$) mesh on germanene. The band structures calculated using hybrid functionals (HSE06) are also presented as Supplement. It should be noted that the band gaps of the semiconductor phases are almost doubled after the HSE corrections.  The dramatic effects of DB coverage is clearly seen in these band structures. While TDP is a nonmagnetic metal, HDP is a nonmagnetic semiconductor with an indirect band gap of 0.53 eV. DHP, which is a metal according to GGA, becomes a narrow band gap semiconductor after HSE06 correction. The ($4\times4$) mesh of DBs is a magnetic, narrow band gap semiconductor. DB gives rise to localized states at $\sim$ -2 eV and -7 eV below the Fermi level.

Incidently, further to Ge DB on germanene, Si adatoms on germanene can also construct asymmetric DB structures, as such that D$_1$ = Si, but D$_2$ = Ge. The formation of this asymmetric DB is also exothermic and occurs spontaneously as long as a Si adatom on germanene is available. These asymmetric DB, as well as their periodic structures on germanene exhibit properties different from host DBs and multiply the functionality of DB based new phases of germanene. For example, the Si-Ge asymmetric dumbbell structure having hexagonal lattice over the ($4 \times 4$) supercell of germanene is a nonmagnetic metal with a binding energy, $E_{b}$=3.86  eV per DB. In contrast, the symmetric germanene-DB having the same lattice is a magnetic semiconductor, and $E_b$=3.44 eV as shown in ~\ref{table1}. We note that like the Si-Ge asymmetric dimer on germanene, also Ge-Si asymmetric dimer can be constructed on silicene spontaneously.

In conclusion, we showed that a Ge adatom constructs DB structure on germanene through an exothermic and spontaneous process. Moreover, through periodically repeating decoration of DBs new phases can grow on germanene. These stable DB based phases exhibit diverse electronic and magnetic structures, which show remarkable changes with the coverage of DBs. In particular, HDP which has two DBs in each ($\sqrt{3} \times \sqrt{3}$) periodic supercell of germanene with a nearest neighbor distance of 3.85 \AA~ attains highest cohesive energy per unit area and forms relatively larger honeycomb pattern over the buckled honeycomb pattern of germanene. It is an indirect band gap semiconductor and appears to be favorable energetically as compared to other phases. The dumbbell based phases of germanene and their multilayer structures bring about new class of single layer materials and can be precursors of 3D layered Ge.

\section{Acknowledgement}
The computational resources have been provided by TUBITAK ULAKBIM, High Performance and Grid Computing Center (TR-Grid e-Infrastructure). SC and VOO acknowledge financial support from the Academy of Sciences of Turkey(TUBA). Authors acknowledge helpful discussions with Dr Seymur Cahangirov, who pointed out DHP.

\section{Supporting Information}
The electronic band structure of the different phases of germanene calculated with hybrid functionals, atomic structures of the various phases of germanene + DB, and phonon dispersion curves for H/FDP and R/RDP structures are available free of charge via the Internet at http://pubs.acs.org.


\begin{thebibliography}{99}

\bibitem{geim}
Geim, A. K.; Novoselov, K. S. The Rise of Graphene.  Nat. Mater. \textbf{2007}, \emph{6}, 183.

\bibitem{novo}
Novoselov, K. S.; Jiang, D.; Booth, T. J.; Khotkevich, V. V.; Morozov, S. V.; Geim, A. K. Two-Dimensional Atomic Crystals. Proc. Natl. Acad. Sci. USA \textbf{2005}, \emph{102}, 10451.

\bibitem{mos2}
Joensen, P.; Frindt, R. F.; Morrison, S. R. Single-Layer $MoS_2$. Mater. Res. Bull. \textbf{1986}, \emph{21}, 457-461.

\bibitem{coleman}
Coleman, J. N.; Lotya, M.; O'Neill, A.; Bergin, S. D.; King, P. J.; Khan, U.; Young, K.; Gaucher, A.; De, S.; Smith, R. J.;  \emph{et al.} Two-Dimensional Nanosheets Produced by Liquid Exfoliation of Layered Materials. Science \textbf{2011}, \emph{331}, 568.

\bibitem{naturenano}
Wang, Q. H.; Kalantar-Zadeh, K.; Kis, A.; Coleman, J. N.; Strano, M. S.  Electronics and Optoelectronics of Two-Dimensional Transition Metal Dichalcogenides. Nat. Nanotech. \textbf{2012}, \emph{7}, 699-712.

\bibitem{nature499}
Geim, A. K.; Grigorieva, V. Van der Waals Heterostructures. Nature \textbf{2013}, \emph{499}, 419-425.

\bibitem{nicolosi}
Nicolosi, V.; Chhowalla, M.; Kanatzidis, M.; Strano, M. S.; Coleman, J. N.; Liquid Exfoliation of Layered Materials.  Science \textbf{2013}, \emph{340}, 6139.

\bibitem{takeda}
Takeda, K.; Shiraishi, K. Theoretical Possibility of Stage Corrugation in Si and Ge Analogues of Graphite. Phys. Rev. B \textbf{1994}, \emph{50}, 14916.

\bibitem{engin}
Durgun, E.; Tongay, S.; Ciraci, S. Silicon and III-V Compound Nanotubes: Structural and Electronic Properties. Phys. Rev. B \textbf{2005}, \emph{72}, 075420.

\bibitem{seymur1}
Cahangirov, S.; Topsakal, M.; Akturk, E.; Sahin, H.; Ciraci, S. First-principles Study of Defects and Adatoms in Silicon Carbide Honeycomb Structures.  Phys. Rev. Lett. \textbf{2009}, \emph{102}, 236804.

\bibitem{ansiklo}
Sahin, H.; Cahangirov, S.; Topsakal, M.; Bekarog�lu, E.; Akturk, E.; Senger, R. T.; Ciraci, S. Monolayer Honeycomb Structures of Group-IV Elements and III-V Binary Compounds: First-principles Calculations. Phys. Rev. B \textbf{2009}, \emph{80}, 155453.

\bibitem{can}
Ataca, C.; Akturk, E.; Sahin H.; Ciraci, S. Adsorption of Carbon Adatoms to Graphene and Its Nanoribbons. J. Appl. Phys. \textbf{2011}, \emph{109}, 013704.

\bibitem{seymur2}
Cahangirov, S.; Topsakal, M.; Ciraci, S. Armchair Nanoribbons of Silicon and Germanium Honeycomb Structures. Phys. Rev. B \textbf{2010}, \emph{81}, 195120.

\bibitem{lelay}
Vogt, P.; De Padova, P.; Quaresima, C.; and Avila, J.; Frantzeskakis, E.; Asensio, M. C.; Resta, A.; Ealet, B.; Le Lay, G. Silicene: Compelling Experimental Evidence for Graphenelike Two-Dimensional Silicon. Phys. Rev. Lett. \textbf{2012}, \emph{108}, 155501.

\bibitem{lelay2}
De Padova, P.; Quaresima, C.; Ottaviani, C.; Sheverdyaeva, P. M.; Moras, P.; Carbone, C.; Topwal, D.; Olivieri, B.; Kara, A.; Oughaddou, H.;  \emph{et al.} Evidence of Graphene-Like Electronic Signature in Silicene Nanoribbons. Appl. Phys. Lett. \textbf{2010}, \emph{96}, 261905.

\bibitem{lelay3}
Padova, P.; et al. Evidence of Dirac fermions in Multilayer Silicene.  Appl. Phys. Lett. \textbf{2013}, \emph{102}, 163106.

\bibitem{germanene}
Li, L.; Lu, S.; Pan, J.; Qin, Z.; Wang, Y.; Wang., Y; Cao, G.; Du, S.; Gao, H. Buckled Germanene Formation on Pt(111). Adv. Mater. \textbf{2014}, DOI:10.1002/adma.201400909.

\bibitem{lelay4}
http://meetings.aps.org/link/BAPS.2014.MAR.T51.7

\bibitem{germanane}
Bianco, E.; Butter, S.; Jiang, S.; Restrepo O. D.; Windi, W.; Goldberg, J. E. Stability and Exfoliation of Germanane: A Germanium Graphene Analogue. ACS Nano \textbf{2013}, \emph{7}, 4414-4421.

\bibitem{grimme06}
Grimme, S. J. Semiempirical GGA-Type Density Functional Constructed with a Long-Range Dispersion Correction. Comput. Chem. \textbf{2006}, \emph{27}, 1787.

\bibitem{blochl94}
Blochl, P. E. Projector Augmented-Wave Method. Phys. Rev. B \textbf{1994}, \emph{50}, 17953.

\bibitem{pbe}
Perdew, J. P.; Burke, K.; Ernzerhof, M. Generalized Gradient Approximation Made Simple. Phys. Rev. Lett. \textbf{1996}, \emph{77}, 3865-3868.

\bibitem{vasp}
Kresse, G.; Furthmuller, J. Efficient Iterative Schemes for Ab-initio Total Energy Calculations Using a Plane-Wave Basis Set. Phys. Rev. B \textbf{1996}, \emph{54}, 11169-11186.

\bibitem{hse}
Heyd, J.; Scuseria, G. E.; Ernzerhof, M. Hybrid Functionals Based on a Screened Coulomb Potential. J. Chem. Phys. \textbf{2003}, \emph{118}, 8207-8215.


\bibitem{alfe2009phon}
Alfe, D.; PHON: A program to calculate phonons using the small displacement method. Comp. Phys. Commun. \textbf{2009}, \emph{180}, 2622-2633.


\bibitem{kittel}
Kittel, C. \textit{Introduction to Solid State Physics}, John Wiley and Sons, New York \textbf{1996}.

\bibitem{socsi}
Liu, C.; Jiang, H.; Yao, Yugui. Low-Energy Effective Hamiltonian Involving Spin-Orbit Coupling in Silicene and Two-Dimensional Germanium and Tin. Phys. Rev. B. \textbf{2011}, \emph{84}, 195430.


\bibitem{ongun1}
\"{O}z\c{c}elik, V. O.; Gurel, H. H.; Ciraci, S. Self-Healing of Vacancy Defects in Single-Layer Graphene and Silicene. Phys. Rev. B \textbf{2013}, \emph{88}, 045440.


\bibitem{kaltsas}
Kaltsas, D.; Tsetseris, L. Stability and Electronic Properties of Ultrathin Films of Silicon and Germanium. Phys. Chem. Chem. Phys. \textbf{2013}, \emph{15}, 9710-9715.


\bibitem{ongun2}
\"{O}z\c{c}elik, V. O.; Ciraci, S. Local Reconstructions of Silicene Induced by Adatoms. J. Phys. Chem. C \textbf{2013}, \emph{49}, 26305-26315.

\bibitem{seymur3}
Cahangirov, S.; Audiffred, M.; Tang, P.; Iacomino, A.; Duan, W.; Merino, G.; Rubio, A. Electronic Structure of Silicene on Ag(111): Strong Hybridization Effects. Phys. Rev. B \textbf{2013}, \emph{88}, 035432.



\end{thebibliography}
\end{document}